\documentstyle[12pt]{article}
\begin{document}

\renewcommand{\thefootnote}{\fnsymbol{footnote}}

\begin{flushright}
LBL-35724
\end{flushright}

\vspace{1cm}
 
\begin{center}
  {\bf Secondary phi meson peak as an indicator of QCD phase transition \\
in ultrarelativistic heavy ion collisions}
\end{center}

\vspace{0.5cm}
\begin{center}
{M. Asakawa$^a$\footnote{Electronic address: yuki@nsdssd.lbl.gov}
and
C. M. Ko$^b$\footnote{Electronic address: ko@comp.tamu.edu}\\
{~}\\}
{\it
$^a$Nuclear Science Division, MS 70A-3307, \\
Lawrence Berkeley Laboratory, Berkeley, CA 94720\\
{~}\\
$^b$Cyclotron Institute and Physics Department\\
Texas A{\&}M University \\
College Station, TX 77843-3366\\
}
\end{center}

%\vspace{4cm}
\vspace{1cm}
\centerline{\bf Abstract}

In a previous paper, we have shown that a double phi peak structure appears
in the dilepton invariant mass spectrum if a first order QCD phase transition
occurs in ultrarelativistic heavy ion collisions. Furthermore, the transition
temperature can be determined from the transverse momentum distribution of the
low mass phi peak. In this work, we extend the study to the case that a smooth
crossover occurs in the quark-gluon plasma to the hadronic matter transition.
We find that the double phi peak structure still exists in the dilepton
spectrum and thus remains a viable signal for the formation of the quark-gluon
plasma in ultrarelativistic heavy ion collisions.
\medskip

\vspace{0.5cm}

\newpage

In a previous paper \cite{ak94}, we have proposed that a double phi peak
structure in the dilepton invariant mass spectrum from ultrarelativistic
heavy ion collisions can be used to confirm the phase transition from the
quark-gluon plasma to the hadronic matter.  Furthermore, the transverse
momentum distribution of the low mass phi peak allows us to determine the
transition temperature.

The low mass phi peak results from the decay of phi mesons with reduced
in-medium mass during the transition. The reduction of the phi meson mass
in hot hadronic matter is a result of the partial restoration of chiral
symmetry \cite{pi82,ha85,ha92}. In normal effective field theory,
hadron masses do not decrease at finite temperature and/or density
\cite{song,hermann}. This result is, however, incomplete, because it neglects
the modification of the vacuum due to medium effects. In QCD, the vacuum can be
described in terms of expectation values of quark and gluon operators, i.e.,
condensates. The correlation function for current operators defined by 
quark fields can be calculated as a sum of condensates in the deep Euclidean
region and can be identified as hadron spectrum function in timelike region.
Values of the correlation function in these two regions are related
to each other through the dispersion relation. The behavior of condensates
at finite temperature and/or density can be calculated with appropriate
approximation. Using these condensates on one side of the dispersion relation
and the hadron spectrum function from an effective theory on the other side,
it has been shown that hadron masses should decrease in the medium in order
to satisfy the dispersion relation \cite{akprr}. In the effective field theory,
the vacuum effect can be also included via the nucleon antinucleon
polarization. Recent studies show that this indeed leads to a reduction of
meson masses at finite temperature and/or density \cite{gao,jpw93,sh94,sxk94}.

If a first order phase transition occurs between the quark-gluon plasma
and the hadronic matter in heavy ion collisions as assumed in Ref. \cite{ak94},
the system spends a relatively long time in the mixed phase (about 10-15 fm)
during which the phi meson mass stays almost constant at a value
different from that in free space. Since the duration of the mixed phase
is not negligible compared to the lifetime of the phi meson in vacuum
($\sim 45$ fm), a low mass phi peak besides the normal one thus appears
in the dilepton spectrum.  As the transverse flow during the mixed phase
is not appreciable, the transverse momentum distribution of the low mass
phi meson is largely determined by the temperature of the mixed phase
and provides thus the information on the transition temperature.

In the scenario described in Ref. \cite{ak94}, we have ignored
the following effects: 
(i) the collisional broadening of the phi meson width due to its interaction
with hadrons; (ii) the increase of the phi meson width in the mixed phase due
to its interaction with partons in the quark-gluon phase; and
(iii) the possibility of a smooth transition from the quark-gluon plasma
to the hadronic matter instead of the mixed phase. If these effects
are large, then the secondary phi peak proposed in Ref. \cite{ak94}
may not appear. Effects (i) and (ii) have recently been studied in
Refs. \cite{sk93} and \cite{sk94}, respectively. It has been found that
the collision of the phi meson with hadrons increases its width to
about 10 MeV while its interaction with partons adds another few
MeV in the width. The resulting phi meson width remains small
enough to make the secondary phi peak visible. In this paper, we shall
study if the double phi peak structure is still present in the dilepton
spectrum in the case of a smooth transition from the quark-gluon plasma
to the hadronic matter.

According to recent lattice calculations \cite{co90}, the QCD phase
transition is a crossover but very close to the first order one. There
exists a sudden change in the entropy density within a temperature
interval of less than ${\sim 10}$ MeV \cite{cr92}. To incorporate these
features of the QCD phase transition, we parametrize the temperature dependence
of the entropy density $s$ of the hot matter as follows:
\begin{eqnarray}
s(T) & = & \frac{ms_h(T)\left (1-\tanh\left(\frac{T-T_c}{\Gamma}\right)\right)
+ns_q(T)\left (1+\tanh\left(\frac{T-T_c}{\Gamma}\right)\right)}
{m\left (1-\tanh\left(\frac{T-T_c}{\Gamma}\right)\right)
+n\left (1+\tanh\left(\frac{T-T_c}{\Gamma}\right)\right)}\nonumber \\
& = & \frac{ms_h(T)\left (1-\tanh\left(\frac{T-T_c}{\Gamma}\right)\right)
+ns_q(T)\left (1+\tanh\left(\frac{T-T_c}{\Gamma}\right)\right)}
{(m+n) + (n-m)\tanh\left(\frac{T-T_c}{\Gamma}\right)},
\label{sdens}
\end{eqnarray}
where $s_h (T) = 12 aT^3$ and $s_q (T) = 148 aT^3$ with
$a={\pi^2}/{90}$ are the bag model entropy density with two flavors in
the hadron and the quark phase, respectively;
$T_c$ is the critical temperature; $m,~n$ and $\Gamma$ are constants.
The typical width of the phase transition is given by $2\Gamma$ with
$\Gamma\sim5$ MeV according to lattice calculations. For temperatures
that satisfy $|T-T_c|\gg\Gamma$, the entropy density given by Eq. (\ref{sdens})
approaches asymptotically to $s_h(T)$ and $s_q(T)$, respectively, for $T$
below and above $T_c$. We have introduced $m$ and $n$ to include
the possibility of an asymmetric phase transition. The case $m/n=$ 1
corresponds to a symmetric phase transition. For $m/n >1 $, the transition
is asymmetric, and the entropy density changes more in the quark phase
than in the hadron phase\footnote{Strictly speaking, if the phase
transition is a crossover, the distinction between the hadron phase and
the quark-gluon phase does not exist. Here we assume, however, that the
system is in the quark-gluon phase if $T\geq T_c $ and the hadron phase
if $T< T_c $.}.

The pressure $P$ and the energy density $e$ can be straightforwardly
evaluated from the entropy density. In this paper, we
take the chemical potential to be zero, as we are interested in the
central region of ultrarelativistic heavy ion collisions, where the baryon
density is expected to almost vanish. Then, the pressure and the energy
density are given by
\begin{eqnarray}
P(T)&=&\int_{0}^{T}s(t)dt, \nonumber\\
e(T)&=&Ts(T)-P(T).
\end{eqnarray}
We note that only one of the three quantities $s$, $P$, and $e$ is
independent at fixed temperature $T$ and chemical potential. In Figs. 1-3,
we show $s$, $P$, and $e$ by solid lines as functions of temperature.
The parameters used in evaluating these quantities are $T_c= 180$ MeV,
$\Gamma= 5$ MeV, and $m/n=1$. The dashed lines show the equation of state
obtained from the standard bag model.

In the following calculations, we use the same temperature dependence of the
phi meson mass as in Ref. \cite{ak94}. This is shown in Fig. 4.
The decrease of the phi meson mass is mainly due to the presence of
a considerable number of strange particles at high temperatures. The details
can be found in Refs. \cite{ak94,ak92}. As in Ref. \cite{ak94}, the
rho meson mass is taken to have the following temperature dependence
\begin{equation}\label{rho}
\frac{m_{\rho} (T)}{ m_{\rho} (T=0)} =
\left [1-\left (\frac{T}{T_c}\right )^2\right ]^{1/6},
\end{equation}
and the omega meson mass is independent of temperature. The difference
in the temperature dependence of vector meson masses is due to the
difference  in their isospin \cite{hkl92}. We would like to point out
that the double phi peak structure in the dilepton spectrum depends more on
the existence of the shift of the phi meson mass at finite temperature
than the magnitude of the shift.

We assume that in ultrarelativistic heavy ion collisions the system has
a cylindrical symmetry and is in thermal equilibrium. All formulas used
in the following calculation are given in Ref. \cite{ak94}. In particular,
we include a normalized smearing function of Gaussian form for the phi
meson mass in order to take into account the experimental mass resolution:
\begin{equation}
F_{\phi} (M, m_\phi (T)) = \frac{1}{\sigma\sqrt{2\pi}} e^{-(M-m_\phi (T))^2
/ 2\sigma^2 },
\end{equation}
where $\sigma$ is a constant and is taken to be 10 MeV \cite{ph92}.

Assuming boost invariance, we have carried out a hydrodynamical calculation
with transverse flow for a central collision of
${\rm ^{197}\!Au +\!{^{197}\!Au}}$. We have modified the code
of Ref. \cite{lm91} to include the smooth quark-gluon plasma to
hadronic transition. We have used the following values for the parameters
\cite{ak94,ak93}:
the initial proper time, $\tau_0=$ 1 fm, and the initial radial velocity at the
surface of the cylinder, $v_0=$ 0; the initial temperature, $T_0 =$ 250 MeV,
the critical temperature, $T_c =$ 180 MeV, and the freeze out temperature,
$T_f =$ 120 MeV; one half of the typical width of the crossover transition,
$\Gamma =$ 5 MeV and the asymmetry factor of the crossover, $m/n=$ 1.
Unless specified otherwise, we use these default values in the calculation.
In addition, we have assumed that the volume fraction of the hadron phase
is 1 and 0 at $T<T_c $ and $T\geq T_c$, respectively.

In Fig. 5, we show by the solid line the invariant mass distribution
of lepton pairs $dN/dMdy$. We see that the second phi peak between
the omega meson and the normal phi meson is still visible as in the case
of a first order phase transition \cite{ak94}. The low mass phi peak is,
however, somewhat broadened as the temperature in the present case
does not stay exactly at the same value during the transition.
For comparison, we have also plotted in Fig. 5 the result for an
asymmetric case by the dashed line.  The asymmetric factor is chosen
to be $m/n = 3$. In this case, the entropy density drops much more
in the quark-gluon phase than in the hadron phase. This makes,
however, practically no difference in the dilepton spectrum as the
temperature stays near the critical temperature $T_c=180$ MeV
(doted line in Fig. 6) for a relatively long time in both the symmetric
and the asymmetric case which are shown in Fig. 6 by the solid and the
dashed line, respectively.  To understand this, we take the Bjorken scaling
solution without transverse expansion and denote the critical entropy
density as $s_c$, which is given by
\begin{equation}
s_c = \frac{m s_h (T_c) + n s_q (T_c)}{m+n} .
\end{equation}
The duration ${\mit\Delta}\tau_c$ for which the temperature remains
near constant is then approximately given by
\begin{eqnarray}
{\mit\Delta}\tau_c & = & \frac{\tau_0 s_q (T_0)}{s_h(T_c)}
-\frac{\tau_0 s_q (T_0)}{s_c} \nonumber \\
& = & \frac{\tau_0 s_q (T_0 )}{s_h (T_c)}
\frac{\frac{s_q (T_c)}{s_h (T_c)} -1}
{\frac{s_q (T_c)}{s_h (T_c)} + \frac{m}{n} }.
\end{eqnarray}
Since ${s_q (T_c)}/{s_h (T_c)}$ is very large ($=37/3$),
${\mit\Delta}\tau_c$ is barely affected by any reasonable
change of the asymmetry factor.

In Ref. \cite{ak94}, we have pointed out that for a first order phase
transition the second phi peak does not appear if the initial temperature is
less than the critical temperature.  This is not the case for a smooth
crossover as shown Fig. 7, in which the initial temperature is taken
to be 175 MeV and is less than the critical temperature. The reason for
this is that for the crossover transition the temperature does not
stay exactly at $T_c$ but around $T_c$, so the second phi peak is expected to
be visible if $T_c -T_0$
{\Large\lower3pt\hbox{$\buildrel <\over{\mbox{\normalsize $\sim$}}$}}
$\Gamma$ is satisfied even if $T_0 < T_c$. The result of our numerical
calculation supports this.

If the critical temperature is small, it is possible that the second
phi peak may not be observed due to the small separation from the
normal phi meson peak. In Fig. 8, we show  by the solid line
the result from the calculation with a lower critical temperature,
$T_c = 160$ MeV. The second peak is still visible. However, if the
width parameter $\Gamma$ of the crossover transition is larger, the second
phi peak becomes broader and merges to the normal one. This is shown by the
dashed line in Fig. 8 for a larger width parameter, $\Gamma= 10$ MeV, and
the same critical temperature $T_c = 160$ MeV. We note that as in a
first order transition our results do not change qualitatively for
higher initial temperatures \cite{ge92,sh92,km92} as long as the
critical temperature is kept the same \cite{ak94}.

In Ref. \cite{ak94}, we have also pointed out that the critical
temperature for the QCD phase transition can be extracted reasonably
accurately from the transverse momentum distribution of the low mass phi meson
peak in the dilepton spectrum. A similar proposal to measure the phase
transition temperature with the transverse momentum distribution of dileptons
from rho mesons was proposed by Seibert \cite{sei92}. The method with the
phi meson, however, has several advantages over the latter one:
(i) The second peak originates exclusively from the matter near the
critical temperature, so the signal to noise ratio is large; (ii) The
transverse flow is still small near the phase transition. Therefore,
the transverse momentum distribution of the second peak reflects more
faithfully the critical temperature; and (iii) The phi peak is narrower
than the rho peak, which makes the subtraction of backgrounds easier.
We have repeated the same procedure as we did in Ref. \cite{ak94} and
have confirmed that the transition temperature can also be extracted
from the low mass phi meson peak in the case of the crossover transition.
In Fig. 9, we show the slope parameter of the dilepton distribution
at small transverse momenta as a function of the initial temperature.
The solid curve is the slope parameter of the low mass peak at about 916 MeV.
It changes only slightly as the initial temperature increases.
On the other hand, the slope parameter of the normal phi meson peak at
1019 MeV changes significantly as the initial temperature increases
due to the resultant development of an appreciable transverse flow in
the hadron phase. We note especially the sudden increase of the slope
parameter of the normal phi meson peak in the region,
$ T_c - \Gamma$
{\Large\lower3pt\hbox{$\buildrel <\over{\mbox{\normalsize $\sim$}}$}}
$T$
{\Large\lower3pt\hbox{$\buildrel <\over{\mbox{\normalsize $\sim$}}$}}
$T_c + \Gamma $, where the entropy density changes rapidly as
the temperature increases.

In summary, due to the reduction of the phi meson mass in a hot matter
and the sudden change of the entropy density at the phase transition,
a distinct low mass peak besides the normal one appears in the
dilepton spectrum even if the quark-gluon plasma to hadronic matter
transition in heavy ion collisions is not first order. This concludes
a series of investigations on the possibility of a double phi peak structure
in the dilepton spectrum from ultrarelativistic heavy ion collisions.
With all factors which could potentially weaken the double phi peak
structure ruled out, the low mass phi peak is thus a credible tool
to verify the occurrence and to determine the critical temperature
of QCD phase transition in future ultrarelativistic heavy ion experiments.

\bigskip

M.A. was supported by the Director, Office of Energy Research,
Office of High Energy and Nuclear Physics, Divisions of High Energy Physics
and Nuclear Physics of the U.S. Department of Energy under Contract
No. DE-AC03-76SF00098. C.M.K. was supported in part by the National
Science Foundation under Grant No. PHY-9212209 and the Welch Foundation under
Grant No. A-1110.

%\bigskip
\newpage
\centerline{\bf Figure Captions}
\vskip 15pt
\begin{description}
\item[Fig. 1] Entropy density as a function of temperature. The number
of flavor is taken to be two and $T_c$ is 180MeV.
Solid and dashed lines correspond, respectively, to the crossover
transition with
$\Gamma=$ 5 MeV and $m/n$= 1, and the first order phase transition with
the bag model equation-of-state.
\item[Fig. 2]
Same as Fig. 1 for the energy density.
\item[Fig. 3]
Same as Fig. 1 for the pressure.
\item[Fig. 4] Temperature-dependent phi meson mass in a hot
hadronic matter.
\item[Fig. 5] Dilepton invariant mass spectrum at central rapidity.
The solid curve is the result from the hydrodynamical calculations with
the default parameter set. The dashed curve is obtained by changing
the asymmetry factor of the phase transition, $m/n$, to 3.
\item[Fig. 6] Temperature as a function of proper time. The solid line
corresponds to the case using the default parameters and the dashed line
is obtained by changing $m/n$ to 3. The dotted line is the critical
temperature used in the calculation.
\item[Fig. 7] Dilepton invariant mass spectrum at central rapidity.
The default parameters are used except that
the initial temperature is $T_0 =$ 175 MeV.
\item[Fig. 8] Dilepton invariant mass spectrum at central rapidity.
For the solid line, the default parameters are used
except that the critical temperature is $T_c =$ 160 MeV.
For the dashed line, in addition, the width parameter of the transition
$\Gamma$ is changed to 10 MeV.
\item[Fig. 9] The slope parameter of the phi meson transverse momentum
distribution as a function of the initial temperature.
Except the initial temperature, the default parameters are used.
Solid and dashed curves correspond to the low mass
and the normal phi meson peak, respectively. The dotted line is the critical
temperature used in the calculation.
\end{description}

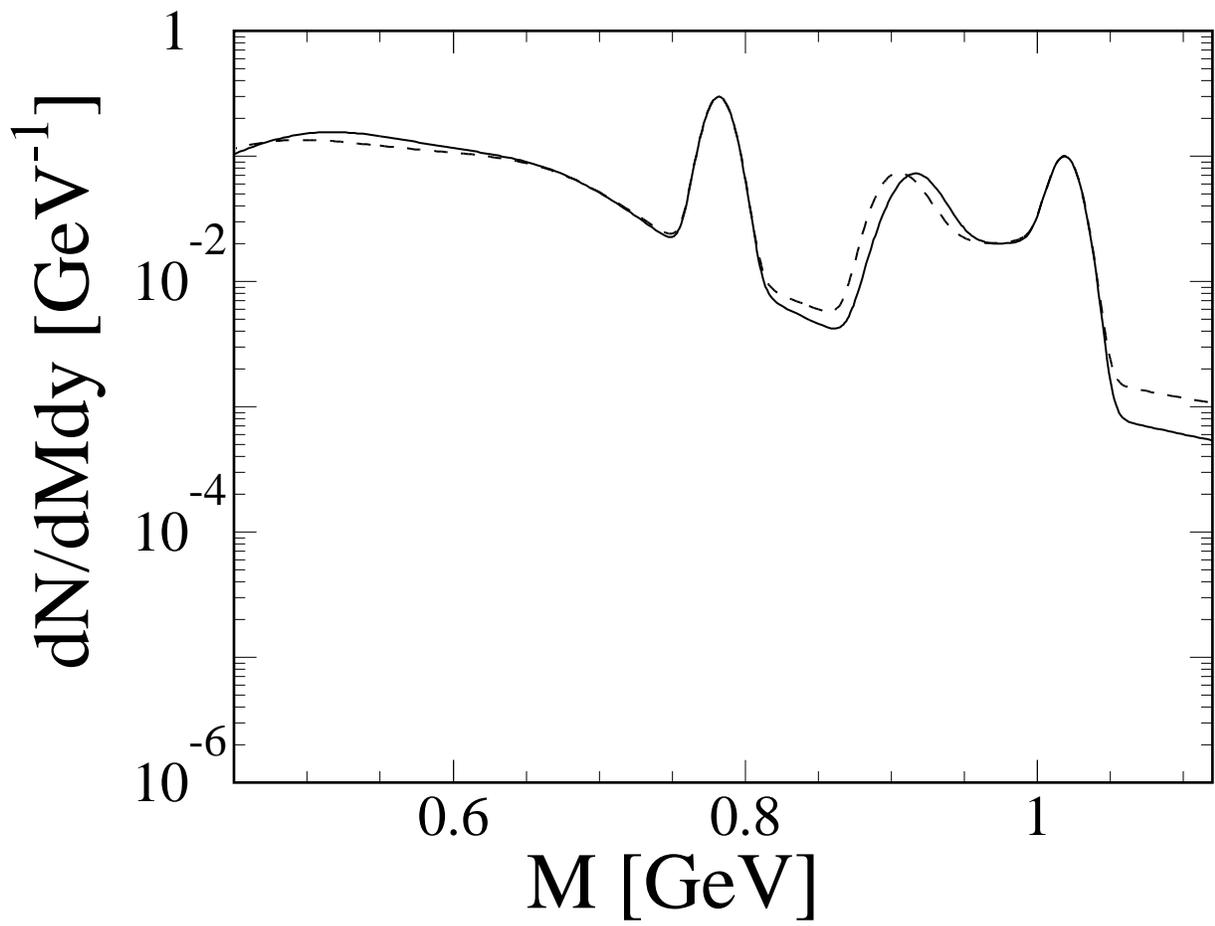
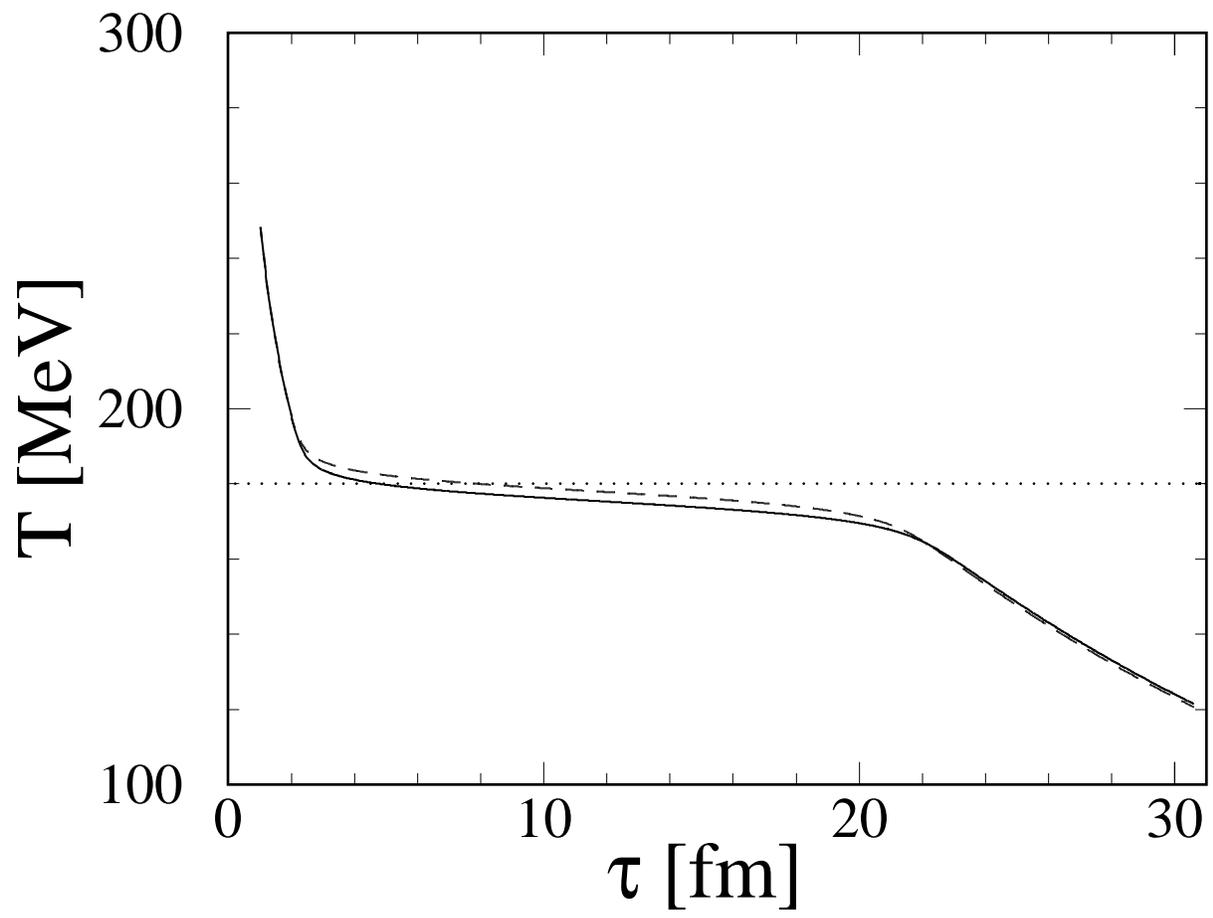
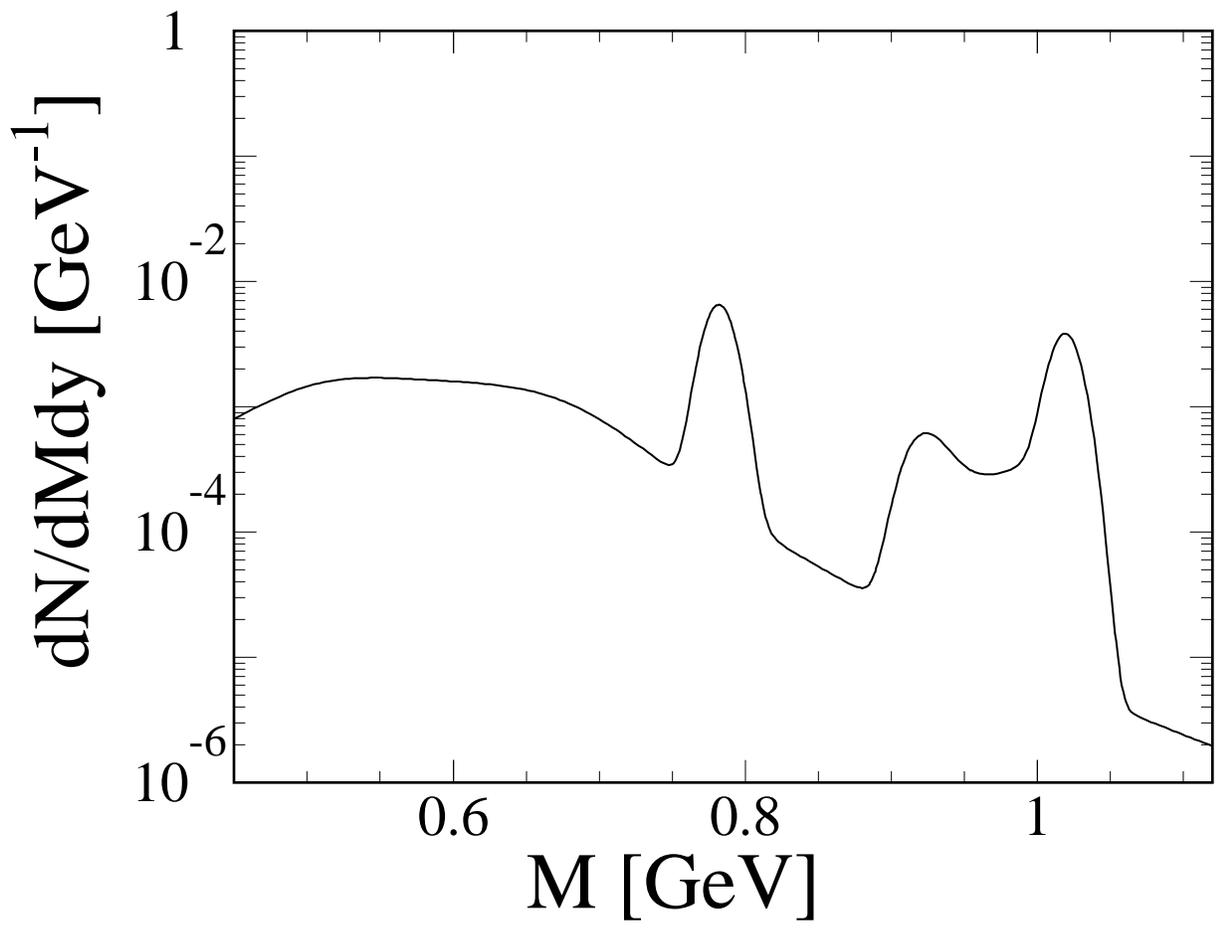
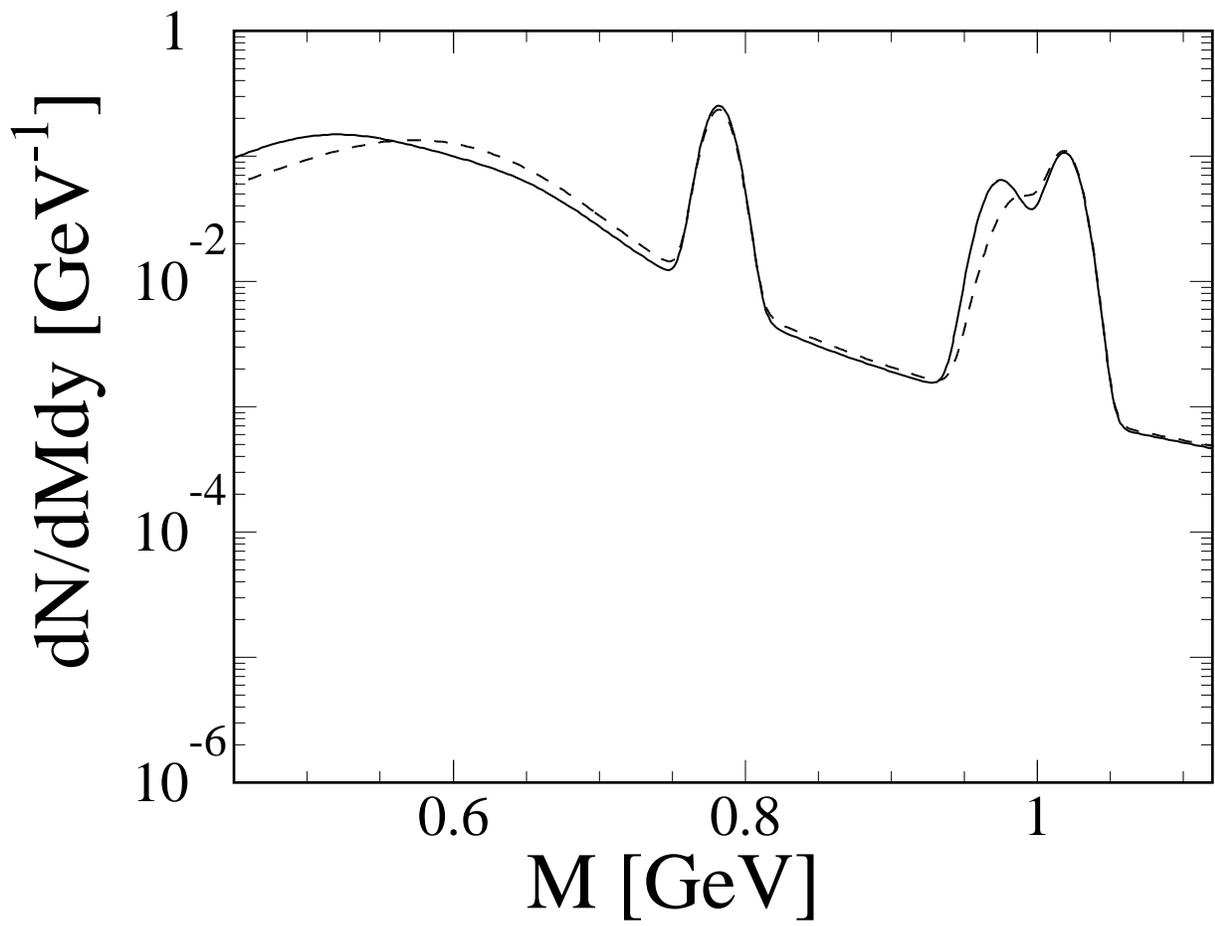
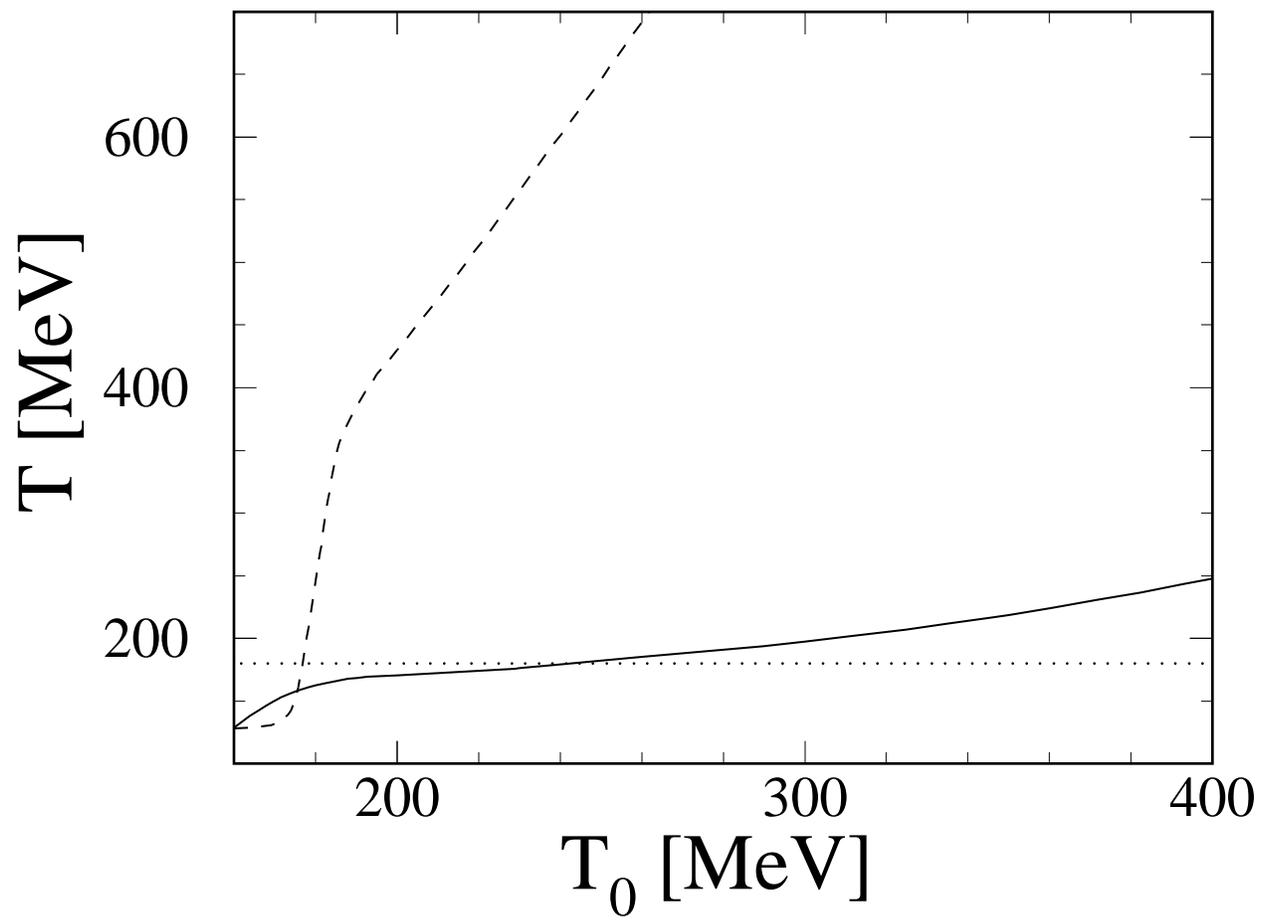
\end{document}